\documentclass[12pt]{article}%
\usepackage{amsmath}
\usepackage{amsfonts}
\usepackage{amssymb}
\usepackage{graphicx}%
\begin{document}
\begin{titlepage}
\begin{flushright}
PUPT-2190\\
hep-th/0512310
\end{flushright}
\vspace{7 mm}
\begin{center}
\huge{Supermagnets and Sigma models}
\end{center}
\vspace{10 mm}
\begin{center}
{\large
A.M.~Polyakov\\
}
\vspace{3mm}
Joseph Henry Laboratories\\
Princeton University\\
Princeton, New Jersey 08544
\end{center}
\vspace{7mm}
\begin{center}
{\large Abstract}
\end{center}
\noindent
We discuss new methods for  non-compact sigma models with and without RR fluxes.The methods include  reduction to one dimensional supermagnets,supercoset constructions and   supertwistors.This work is a first step towards the solution of these models,which are important in several areas of physics.
I dedicate it to the memory of Volodya Gribov.
\vspace{7mm}
\begin{flushleft}
December 2005
\end{flushleft}
\end{titlepage}

\section{\bigskip\ Introduction}

Nonlinear sigma models describe low energy interaction of  Goldstone particles
and are important in many parts of modern physics. The models in two
dimensions are especially interesting since the infrared effects in this case
are highly nontrivial. Such models are relevant in string theory, quantum Hall
effect, spin chains, disordered systems etc.

It has been shown long ago that in general the models with the nonabelian
symmetry groups acquire a non-perturbative mass gap [ 1]. This result
contradicted to  the well known fact that in the Heisenberg antiferromagnet
(which has O(3) symmetry) spin waves are gapless. The puzzle was resolved
seven years later [2,3] when it was argued that the Heisenberg model with the
half integer spin must be described by the sigma model with the fine tuned
topological term, which makes it gapless. In the case of integer spin, the gap
found in [1] is expected to appear (and does so experimentally). At the same
time, even for the integer spins, there are special forms of the interaction
leading to a completely integrable gapless antiferromagnets. So, the Goldstone
particles can remain massless when they want to.

In this paper I will begin developing new methods for analysis and
interpretation of the massless sigma models. My refreshed interest in the old
toy is related to the gauge/strings duality, the key elements of which are the
supersymmetric sigma models. Existing methods are painfully inadequate for
getting physical information in this case as well as in the other areas. The
purpose of this paper is to outline new ways and hopes for their solution.

Let us begin with a brief summary of the Haldane-Affleck approach. One
considers a Heisenberg antiferromagnet. Its hamiltonian has the form%
\begin{equation}
\mathcal{H=}\sum S_{x}S_{x+1}%
\end{equation}
where $S$ are spin one half matrices. They can be expressed in terms of free
fermions, $a_{x\alpha},\alpha=1,2$ as%
\begin{align}
S_{x}  &  =a_{x}^{+}\sigma a_{x}\\
a_{x}^{+}a_{x}  &  =1
\end{align}
where $\sigma$ are the Pauli matrices and the constraint (3) ensures that
there is exactly one fermion at each site (half- filled band). One then can
give a plausible argument that in the continuous limit the fermions become
relativistic, left and right moving Dirac particles with isotopic spin,
$\psi_{L\alpha}(x)$ and $\psi_{R\alpha}(x)$. This doubling of the fermions
occurs since the continuous limit enforces the continuity of the even and odd
sites separately. The Lagrangian for these fields has the standard Dirac form%
\begin{equation}
L=\psi^{+}\overset{\wedge}{(\partial}+\overset{\wedge}{A})\psi
\end{equation}
Here the $U(1)$ gauge field $A$ is a Lagrange multiplier ensuring the
constraint ( 3). The spin field $S_{x}$ has the following behavior in the
continuos limit (reflecting its antiferromagnetic nature)%
\begin{equation}
S_{x}=(-1)^{x}n(x)+l(x)
\end{equation}
where $l\ll n$ are continuous functions and in terms of the Dirac fermions
\begin{equation}
n(x)=\psi_{L}^{+}\sigma\psi_{R}+c.c
\end{equation}
Without $A$ field Dirac fermions generate the $SU(2)$ currents $J_{L,R}%
=\psi_{L,R}^{+}\sigma\psi_{L,R}$ which are holomorphic and antiholomorphic and
the $U(1)$ current. The latter is killed by the $A$ integration. The resulting
system is equivalent to the WZNW action (non-abelian bosonization [4] ). For
higher spins one considers several flavors of fermions and gauge out the
flavor currents.

In order to relate these facts to the $O(3)$ sigma model one needs another
representation of the spin operators, coming from the geometrical (Kirillov)
quantization. It is based on the formula
\begin{equation}
\int Dne^{-W(n)}n_{a_{1}}...n_{a_{k}}=Tr(S_{a_{1}}...S_{a_{k}})
\end{equation}
On the right side of this formula we have a trace of the spin matrices in a
representation with the spin $S$ , while on the left side we have an integral
over the unit sphere, which is an orbit of the rotation group. The action has
the form%
\begin{equation}
W(n)=ik\int A(n)dt
\end{equation}
with $A(n)$ being a vector potential of the magnetic monopole and $k=2S$. The
key feature of this action which ensures (7 ) is its variation under the group
transformation%
\begin{align}
\delta n  &  =\omega\times n\\
\delta W  &  =\int(n\frac{d\omega}{dt})dt
\end{align}
With these formulae the geometric quantization follows from the standard Ward
identities, as we will discuss later in the general case. We can now replace
the quantum spins by the classical orbit. Due to the antiferromagnetic order,
we expect in the continuum limit%
\begin{equation}
S_{x}\equiv(-1)^{x}N_{x}\Rightarrow(-1)^{x}n(x)+l(x)
\end{equation}
where $n(x)\gg l(x)$ are continuous functions. The action for our magnet in
this variables takes the form%
\begin{equation}
F=\int dt\{%
%TCIMACRO{\dsum \limits_{x}}%
%BeginExpansion
{\displaystyle\sum\limits_{x}}
%EndExpansion
(-1)^{x}W(N_{x})+%
%TCIMACRO{\dsum \limits_{x}}%
%BeginExpansion
{\displaystyle\sum\limits_{x}}
%EndExpansion
(N_{x}-N_{x+1})^{2}+const\}
\end{equation}
In the continuous limit we should keep only such terms that the sign factor
$(-1)^{x}$ disappears, otherwise the oscillations would kill the sum. That
means that only the terms linear in $l$(x) must appear in the first term of
(12 ) . Using the variation formula (10 ) we get%
\begin{equation}
F\simeq ik\int dt\int dxl(x)[n\times\frac{dn}{dt}]+\int dtdx\{(\nabla
n+2l)^{2}+l^{2}\}
\end{equation}
Integrating out the field $l(x)$ we obtain the result of [2,3 ] - the n-field
sigma model with the theta term at $\vartheta=2\pi S.$

The reason why we need an antiferromagnet to describe a relativistic theory is
the following. Spin changes sign under the time reversal. Hence for a
ferromagnet the presence of the spontaneous magnetization implies breaking of
the CPT-symmetry and the spectrum of the Goldstone particles is
non-relativistic, $\omega=k^{2}.$ In antiferromagnets we can combine
CPT-reflection with the shift $x$ to $x+1.$ This symmetry $\omega
\Rightarrow-\omega$ is not broken and implies the relativistic spectrum
$\omega^{2}=k^{2}$ in the long wave limit. The sigma model is simply the most
general effective lagrangian describing relativistic, $O(3)$ invariant spin waves.

\section{ General groups}

Let us now generalize these considerations for arbitrary groups and
supergroups. The first step is well known - it is Kirillov's theory. This
theory allows to replace traces in a certain representation by integration
over a particular orbit. To set the stage I rederive it by using rather
pedestrian methods. The action has the form%
\begin{equation}
W=\int dtTr(K\Omega^{-1}\frac{d\Omega}{dt})
\end{equation}
where everything is in the adjoint representation. The matrix $K$ depends on
the representation which we eventually want to describe . If this
representation has the highest weight $\mu=(\mu_{1}...\mu_{r})$ ( $r$ is the
rank of the group) and if we have Cartan's generators $H_{k}$ , then
$K=(\mu_{k}H_{k}).$If we consider the variation of $\Omega$ by right
multiplication, $\delta\Omega=\varepsilon\Omega,$where $\varepsilon
=\varepsilon^{A}\tau^{A}$ with $\tau^{A}$ the set of adjoint generators, it
follows from (14 ) that
\begin{equation}
\delta W=\int dtT^{A}\frac{d\varepsilon^{A}}{dt}%
\end{equation}
where $T^{A}=Tr(K\Omega^{-1}\tau^{A}\Omega).$These quantities define an orbit
of the group, since they are invariant (as well as the action) under the
following gauge transformation%
\begin{align}
\Omega &  \Rightarrow\Omega h\\
hKh^{-1}  &  =K
\end{align}
The last equation explicitly shows dependence of the stability group on the
highest weight. Indeed, if we look at the infinitesimal transformations then a
generator defined by a root $\alpha$ belongs to the gauge group if ($\mu
\alpha)=0.$When $\mu$ is a fundamental weight there are non-trivial solutions
of this equation, some of which will be discussed below. For a generic weight
the stability group is generated by the Cartan subalgebra.

In order to derive the Ward identities we notice that under the above right
multiplication we have a transformation law for $T^{A}$%
\begin{equation}
\delta T^{A}=f^{ABC}\varepsilon^{B}T^{C}%
\end{equation}
(here $f^{ABC}$ are the structure constants). Combining (15 ) and (18 ) we get
the equations of motion%
\begin{equation}
\frac{d}{dt}\langle T^{A}(t)T^{B_{1}}(t_{1})...T^{B_{n}}(t_{n})\rangle=%
%TCIMACRO{\dsum \limits_{k}}%
%BeginExpansion
{\displaystyle\sum\limits_{k}}
%EndExpansion
\delta(t-t_{k})f^{AB_{k}C}\langle T^{B_{1}}(t_{1})...T^{C}(t_{k})...T^{B_{n}%
}(t_{n})\rangle
\end{equation}
This is precisely the equation satisfied by the trace of the time-ordered
product of matrices of a given representation, and hence the characters are
given by the functional integrals with the action $W.$ In other words, the
matrices of representation can be replaced by the classical quantities $T^{A}$
defining the orbit. So far we described the well known results in a slightly
unusual way. Now we pass to something new.

\section{Collectivization}

We use this infamous word to describe formation of spin waves out of
fluctuations of individual spins.

Consider a magnet with the Hamiltonian (1) ,except this time $S_{x}^{A}$ are
the matrices in a given representation of a given group. By the same logic as
above we can write the action in the form%
\begin{equation}
F=i%
%TCIMACRO{\dsum \limits_{x}}%
%BeginExpansion
{\displaystyle\sum\limits_{x}}
%EndExpansion
(-1)^{x}W[\Omega_{x}]+(T_{x}^{A}-T_{x+1}^{A})^{2}%
\end{equation}
The factor $(-1)^{x}$ tells us that the representations we use for the
even/odd sites must be conjugate to each other.

Once again we assume that $\Omega_{x}\simeq(1+(-1)^{x}l(x))\Omega(x)$ and that
the adjoint matrix $l(x)$ $\ll1.$It is convenient to work with the matrices in
the adjoint representation, defined as $T(x)=T^{A}\tau^{A}.$Using the formulae
for variations, given above, we get (again keeping only the term in which
$(-1)^{x}$ cancels)%
\begin{equation}
F=\int dtdxTr\{(l\partial_{t}T)+(\partial_{x}T+[l,T])^{2}+[l,T]^{2}\}
\end{equation}
Excluding $l$ we get the equation%
\begin{equation}
\lbrack T,\partial_{x}T]+[T,[T,l]]=2\partial_{t}T
\end{equation}
Since according to our definition $T=\Omega K\Omega^{-1}$ ,its derivative
satisfies%
\begin{align}
\partial_{\alpha}T  &  =[B_{\alpha},T]\\
B_{\alpha}  &  =\partial_{\alpha}\Omega\Omega^{-1}%
\end{align}
If we set $l=B_{1}+\omega,$ we get
\begin{equation}
\lbrack T,[T,\omega]]=2[B_{0},T]=2\partial_{t}T
\end{equation}
The theta term $Q$, which contains both $x$ and $t$ derivatives has the form%
\begin{align}
Q  &  \sim\int dtdxTr(T[B_{0},B_{1}])=\int dtdxTr(K[A_{0},A_{1}])\\
A_{\alpha}  &  =\Omega^{-1}\partial_{\alpha}\Omega
\end{align}
The equaton for $\omega$ can be solved by "rotating" it back to the form%
\begin{align}
\lbrack K,[K,\widetilde{\omega}]  &  =-2[K,A_{0}]\\
\widetilde{\omega}  &  =\Omega^{-1}\omega\Omega
\end{align}
From this we conclude that%
\begin{equation}
\lbrack K,\widetilde{\omega}]=-2A_{0}+h
\end{equation}
where $h$ is a connection in the stability group, $[K,h]=0.$Excluding $h$ from
the action and performing some trivial rescalings, we arrive at the sigma
model action, describing the above antiferromagnet%
\begin{equation}
S=\int dtdx\{Tr[K,A_{\alpha}]^{2}+i\pi\varepsilon_{\alpha\beta}Tr(K[A_{\alpha
},A_{\beta}])\}
\end{equation}
It must be noticed that in general this action is deformed by renormalization,
since nothing prevents functions of $T$ to appear under the trace (since $T$
has classical dimension zero and is invariant under the gauge group). Also,
they may and should be present in the original hamiltonian. However, from the
definition of $T$ it follows that it satisfies polynomial relations, defining
the orbit of the form $TrT^{k}=C_{k}(\mu)$ , where $C_{k}(\mu)$ is related to
the value of the Casimir operators in the given representation. That reduces
or eliminates the number of possible structures. A general invariant
Lagrangian should have the form%
\begin{equation}
L=Tr\{a(T)(\partial_{\alpha}T)^{2}+\varepsilon_{\alpha\beta}b(T)\partial
_{\alpha}T\partial_{\beta}T\}
\end{equation}
The second, parity violating term, is not in general a topological invariant.
However, in some cases it is. For example, for the $O(3)$ model, the
constraint is $T^{2}=const$ and thus $b(T)\sim T$ ; it is easy to check that
we get the standard theta term this way. However, for a general constraint the
second term in (32 ) is parity violating but not topological. The parity
violating terms (which we will call $\varepsilon-$terms) can be easily written
in terms of the left-invariant connections (in a way invariant under the gauge
group $H$ )%
\begin{equation}
L_{\varepsilon}=\varepsilon_{\alpha\beta}Tr(b(K)[K,A_{\alpha}][K,A_{\beta}])
\end{equation}
In many instances this structure degenerates into a topological term, but in
some important cases it doesn't.

Let us give a few less trivial examples of the sigma models related to
antiferromagnets. Take first a vector representation of $SO(D)$, $D=2n,$ which
is described by the fundamental weight corresponding to the first node of the
Dynkin diagram. If we denote the generators by $M_{\mu\nu}$ , for this
representation $K=M_{12}.$The generators commuting with $K$ are $M_{12}$
itself, as well as those of remaining $SO(D-2).$ Hence $T\in\frac
{SO(D)}{SO(D-2)\times SO(2)}$ which is the Grassman manifold. It is
conveniently described by the field of the antisymmetric tensor $t_{\mu\nu},$
satisfying conditions
\begin{equation}
t_{\mu\nu}^{2}=1;t\wedge t=0
\end{equation}
The corresponding sigma model with the theta term has the Lagrangian%
\begin{equation}
L=\frac{1}{2\gamma}(\partial_{\alpha}t_{\mu\nu})^{2}+i\vartheta\varepsilon
_{\alpha\beta}t_{\mu\nu}\partial_{\alpha}t_{\mu\lambda}\partial_{\beta
}t_{\lambda\nu}%
\end{equation}
The claim is that this model describes low energy properties of the quantum
$O(D)$ antiferromagnet in the vector representation. We will see below that
,while being asymptotically free , it has a conformal fixed point described by
free fermions.

Another interesting case is the spinor representation of $O(2n).$For this
representation $K=\frac{1}{2}(M_{12}+M_{34}+...M_{2n-1,2n}).$The stability
group is $U(n)$ and we are dealing with the coset $\frac{SO(2n)}{U(n)}.$This
is the orbit of pure spinors ,which are the Weyl spinors $\lambda$ with the
constraints , analogous to (34)\ , $\overline{\lambda}\gamma_{\mu_{1...\mu
_{p}}}\lambda=0$ for $p<n.$ Again, the conformal point is described by the
Lagrangian for pure spinors and their conjugates and is closely related to the
description of the Berkovits string.

Finally let us discuss supergroups, beginning with the $OSp(1|2).$Its
generators are obtained by adding to $SL(2)$ two spinorial supercharges
$Q_{\alpha}$ .The matrix $\Omega$ can be conveniently written as
$\Omega=e^{(\vartheta Q)}g,$where $g$ is the matrix of $SL(2)$ and
$\vartheta-s$ are two Grassman variables. It is quite easy to find the action
$W(n,\vartheta).$For that we notice that
\begin{align}
\Omega^{-1}d\Omega &  =g^{-1}dg+g^{-1}e^{-(\vartheta Q)}de^{(\vartheta Q)}g\\
e^{-(\vartheta Q)}de^{(\vartheta Q)}  &  =(Qd\vartheta)+\frac{1}{2}\tau
^{a}\overline{\vartheta}\gamma^{a}d\vartheta
\end{align}
Using these formulae we get%
\begin{equation}
W(n,\vartheta)=\int dt\{A(n)+\frac{1}{2}\overline{\vartheta}(\gamma^{a}%
n^{a})\frac{d\vartheta}{dt}\}
\end{equation}
This action performs geometrical quantization of $OSp(1|2),$provided that we
identify the generators as follow%
\begin{align}
S^{a}  &  \Rightarrow(1-\frac{1}{2}(\overline{\vartheta}\vartheta))n^{a}\equiv
N^{a}\\
Q^{\alpha}  &  \Rightarrow\vartheta^{\alpha}\\
(N^{a})^{2}+\overline{\vartheta}\vartheta &  =1
\end{align}

\section{Non-compact supermagnet}

Let us describe the magnet and the sigma model of the last example. The
resulting theory is a closed cousin of the AdS/CFT sigma models. There are two
complications to be overcome. The first is non-compactness of the orbit and
the second is the presence of the grassman dimensions. Let us discuss the
first problem first. Non-compact magnets have been considered before in the
important work [5 ],but we need a somewhat different perspective.

We are looking for a magnet, the hamiltonian of which acts in the Hilbert
(positive norm) space. This is possible if the spin operator at each site acts
in a unitary representation of $SL(2)=Sp(2).$Such representations are
necessarily infinite dimensional. If we perform the geometric quantization
with the phase factor $\exp\pm i\kappa\int A(n)dt$ with real $\kappa$ the
unitarity is to be expected and we have to identify the representation.
Parametrizing the unit vector $n$ as $n=(\cosh\theta,\sinh\theta\cos
\varphi,\sinh\theta\sin\varphi),$we find that the phase factor is equal to
$\exp\pm i\kappa\int\cosh\theta\frac{d\varphi}{dt}dt$ (in the Poincare
coordinates it is even simpler, $\exp\pm i\kappa\int\frac{dx}{dt}\frac{dt}{y}%
$). This corresponds to the unitary representations with the lowest (highest)
weights ,described by the upper (lower) sheet hyperboloids. If $\kappa$ is an
integer, this is a discrete series of representation. Otherwise, this is a
representation of the covering group of $SL(2).$It is useful for our purposes
to construct these representations explicitly as the Verma modulus. Namely,
consider the Virasoro generators ($L_{\pm1},L_{0})$ which form the $SL(2)$
subalgebra. The representation in question is defined by the lowest weight
vector
%TCIMACRO{\TEXTsymbol{\vert}}%
%BeginExpansion
$\vert$%
%EndExpansion
$\kappa\rangle$ satisfying $L_{1}|\kappa\rangle=0$ and $L_{0}|\kappa
\rangle=\kappa|\kappa\rangle.$All other states in the representation are
simply $|\kappa+m\rangle=(L_{-1})^{m}|\kappa\rangle.$This representation can
be obtain from the finite dimensional $SU(2)$ representations by means of
analytic continuation, setting the spin $S=-\kappa.$All these facts are easy
to check by solving the Schrodinger equation for a particle on a hyperboloid
with the phase factor and concentrating on the lowest energy level. The wave
function in this case is proportional to the Jacobi function $P_{m\kappa
}(\cosh\theta),$ and the limit of geometric quantization, when only the phase
factor is left in the lagrangian, corresponds to taking $l=k,$ $m\geq\kappa.$

It is well known that the spin chains of arbitrary integer spin can be made
completely integrable by the special choice of the hamiltonian [6,7] ,which
has the form%
\begin{equation}
\mathcal{H=}\sum_{x}f(S_{x}\cdot\widetilde{S}_{x+1})
\end{equation}
where the function $f$ is expressed in terms of the derivative of the gamma
function. In our case it should require a special consideration. The reason is
that as a result of the factor $(-1)^{x}$ in (20 ), representations at the
even and odd points are conjugate (denoted by tilde) to each other. The
conjugate representation is generated from the highest weight state
$|\widetilde{\kappa}\rangle$ satisfying $L_{-1}|\widetilde{\kappa}\rangle=0$
and $L_{0}|\widetilde{\kappa}\rangle=-\kappa|\widetilde{\kappa}\rangle.$
Integrability is not affected by the above analytic continuation to negative
spins. Indeed the Lax operator has the standard form
\begin{equation}
\mathcal{L}=\lambda+i\overrightarrow{\sigma}\overrightarrow{S}%
\end{equation}
where $\sigma$ are the Pauli matrices and $S$ are the spin operators in any
representation. The fact that at the even points we have the lowest weight
representation and at the odd ones the highest weight, doesn't prevent us from
constructing integrals of motion generated by the $Tr%
%TCIMACRO{\dprod }%
%BeginExpansion
{\displaystyle\prod}
%EndExpansion
\mathcal{L}_{x}(\lambda)$. But constructing the hamiltonian, the Bethe ansatz
and finding the physical vacuum, the excitation spectrum and the central
charge remains an open problem. The technical difficulty is that the product
of representations in (42) has neither highest nor lowest weight. Because of
that we will discuss an alternative approach to Bethe ansatz below.

There is no guarantee that this system has a relativistic limit, but if it
does, this magnet describes (in the infrared limit ) conformal points of the
non-compact $n$- field. These conformal points (if they exist ) must be
unusual. In the compact case of $SU(2)$ we know that such points are described
by the WZNW action and contain holomorphic and antiholomorphic $SU(2)\times
SU(2)$ currents. This can't be the case in the above model since the
non-compact version of current algebra contains negative norm states, while
the lattice model is unitary. Moreover, in the perturbation theory this
hyperbolic $n$ -field is obtained from the standard spherical $n$- field by
changing the sign of the coupling constant. That reverses the asymptotic
freedom to the "zero charge" (IR fixed point at zero coupling). That means
that the non-trivial IR limit can arise only as a result of fine tuning of
higher relevant operators.

Let us generalize these considerations for the $OSp(1|2)$ group. Again it is
convenient to consider it as a subgroup of the super-Virasoro algebra. The
generators are simply $(L_{\pm1},L_{0},G_{\pm\frac{1}{2}}).$The lowest weight
unitary representations are obtained by the action of the raising operator
$G_{-\frac{1}{2}}$ on the state $|\kappa\rangle$ , annihilated by $G_{\frac
{1}{2}}.$The conjugate, highest weight, representations are defined as before.
If we rename these generators by $(S^{a},q^{\alpha}),$with $\alpha$ being a
spinor index and $q$ a Majorana spinor, we can consider an integrable
supermagnet in the form
\begin{equation}
\mathcal{H=}\sum_{x}f(S_{x}\widetilde{S}_{x+1}+\overline{q}_{x}\widetilde
{q}_{x+1})
\end{equation}
Once again the $R$ -matrix for the $OSp(1|2)$ system has been found long ago
[8 ] , but the spectrum of the above (non-compact ) antiferromagnet is not
known. Let us find the corresponding sigma model. As clear from the above
discussion, it contains a hyperbolic unit vector $n$ and its superpartner
$\vartheta.$ It is convenient to write it first in terms of connections. The
corresponding coset space is $OSp(1|2)/O(2).$The connections of $OSp(1|2)$ can
be written down as \{ $B_{\mu}^{a},A_{\mu},\psi_{\mu}^{\alpha}$\} where
$a=1,2$ is the "vertical" index, $A_{\mu}$ is the $SO(2)$ connection and
$\psi_{\mu}^{\alpha}$ are the two spinor connections . They satisfy the
standard Maurer -Cartan equations. The lagrangian must have $SO(2)$ gauge
symmetry. The most general form of it is
\begin{equation}
\mathcal{L=}\frac{1}{2\alpha}\{(B_{\mu}^{a})^{2}+c_{1}\overline{\psi}_{\mu
}\psi_{\mu}+c_{2}\varepsilon_{\mu\nu}\overline{\psi}_{\mu}\gamma_{3}\psi_{\nu
}\}
\end{equation}
where $\alpha$ is a coupling constant; we don't include here the standard
theta term because in the non-compact case it is trivial, but we have instead
the epsilon term, which is not a total divergence. These models appeared in
[9,10,11 ] in connection with gauge/strings duality. They describe the
$AdS_{2}$ space with the RR fluxes. As was shown in [11 ] , they are a part of
the much more interesting cosets , $\frac{OSp(2|4)}{SO(4)\times SO(2)}%
\rightarrow AdS_{4},\frac{SU(4|2)}{SO(5)\times SU(2)}\rightarrow AdS_{5}\times
S^{1},$and $\frac{PSU(4|4)}{SO(5)\times SO(5)}\rightarrow AdS_{5}\times
S^{5}.$In all these cases the lagrangians are basically the same, except the
the type of spinors are different, the matrix $\gamma_{3}$ is replaced by
$\gamma_{5}$ etc. There exists a supermagnet representation for these models.
In order to have conformal symmetries it is necessary to adjust the couplings
$c_{1}$ and $c_{2}$. These conformal points are completely integrable, at
least classically. For the action (45) the Lax representation is as following
[11 ]%
\begin{align}
L_{+} &  =\partial_{+}+A_{+}T_{3}+\lambda^{-1}B_{+}^{a}T_{a}+(\lambda
^{-\frac{1}{2}}\psi_{L+}^{\alpha}+\lambda^{\frac{1}{2}}\psi_{R+}^{\alpha
})Q_{\alpha}\\
L_{-} &  =\partial_{-}+A_{-}T_{3}+\lambda B_{-}^{a}T_{a}+(\lambda^{-\frac
{1}{2}}\psi_{L-}^{\alpha}+\lambda^{\frac{1}{2}}\psi_{R-}^{\alpha})Q_{\alpha}%
\end{align}
where $(T_{3},T_{a},Q_{\alpha})$ are the generators of $OSp(1|2),$and
$\psi_{L,R}=\frac{1\pm\gamma_{3}}{2}\psi.$ As was checked in [11 ] the
equations of motion following from $[L_{+},L_{-}]=0$ correspond to the
lagrangian (45 ) with $c_{1}=0$ and $c_{2}=1.$Amazingly, at the same point the
$\kappa-$symmetry appears and the perturbative beta function is zero. We can
add to this an assumption that at the same point the sigma model becomes
equivalent to the supermagnet (44 ). The Lax representation for the last of
the above cosets has been found in [12]. An interesting different class of
cosets was found in [20]. It deals with the different manifolds, but more
importantly with the cases $c_{1}=c_{2}=1$ . Perhaps the $c_{1}$ term,
together with the added pure spinor lagrangian, can be looked at as fixing of
the $\kappa$ gauge symmetry, with the pure spinors playing the role of the
corresponding ghosts.

\section{ Null vectors and linear sigma models}

The above approach used the discretized theory. It is possible to avoid it.
Let us again consider the $n$ -field first. It is equivalent to the WZNW model
at level one. On the other hand, the same conformal point must appear in the
linear sigma model of the vector field $\overrightarrow{\phi}$ , satisfying
the equation of motion [12]
\begin{equation}
\partial^{2}\overrightarrow{\phi}=\lambda\phi^{2}\overrightarrow{\phi}%
\end{equation}
where $\lambda$ is a coupling constant and the physical mass of the $\phi$
field is tuned to zero. The reason for this "linearization" is that the
constraint $n^{2}=1$ is relaxed by the infrared fluctuations. Whether this
relaxation is relevant , depends on the infrared dynamics.

Let us "solve" this equation using the OPE of the WZNW model ( similar
approach was used in [13 ] for the Ising and other minimal models). It is
based on the fact, understood at the very beginning of conformal field theory,
that the equations of motion must be read from the right to the left,
replacing the products of fields by their OPE.

Let $g$ be an $SU(2)$ matrix entering the action and try to identify
$\overrightarrow{\phi}\sim Tr(\overrightarrow{\sigma}g).$ First of all, in
WZNW model at $k=1$ we have the fusion rule [14 ]%
\begin{equation}
g\times g\times g=[g]
\end{equation}
where brackets mean all possible operators obtained by the action of the
Kac-Moody currents $J_{-n}^{a}$ on the field $g.$ So, the first conclusion is
that the RHS of (48) belongs to the current block%
\begin{equation}
\phi^{2}\phi^{a}=c_{1}\phi^{a}+c_{2}^{abcd}J_{-1}^{b}\overline{J}_{-1}^{c}%
\phi^{d}+c_{3}L_{-1}\overline{L}_{-1}\phi^{a}+...
\end{equation}
where we dropped the higher order terms.

The third term in this equation is just what we need for the (48). The first
term is the physical mass of the $\phi$- field and must be eliminated by hands
( in the language of critical phenomena that corresponds to setting the
temperature to be critical). So, let us concentrate on the second term. We
will show that at $k=1,$ due to the structure of the null vectors, it reduces
to the third term. Consider the states of the form%
\begin{equation}
|\psi\rangle=J_{-1}^{a}|\alpha\rangle
\end{equation}
where
%TCIMACRO{\TEXTsymbol{\vert}}%
%BeginExpansion
$\vert$%
%EndExpansion
$\alpha\rangle$ is a state with spin 1/2 and the projection of the spin equal
to $\pm\frac{1}{2}.$ The state
%TCIMACRO{\TEXTsymbol{\vert}}%
%BeginExpansion
$\vert$%
%EndExpansion
$\psi\rangle$ can be decomposed into spin $\frac{3}{2}$ and spin $\frac{1}{2}$
components. It is well known that the Kac - Moody algebras at the level $k$
have the null vector of the form%
\begin{equation}
(J_{-1}^{+})^{k-2j+1}|j,j\rangle=0
\end{equation}
where the current acts on the highest weight state with the spin $j.$ By
choosing $j=\frac{1}{2}$ and $k=1$ , we conclude that the spin $\frac{3}{2}$
part of the state $|\psi\rangle$ is actually zero. As for the remaining spin
$\frac{1}{2}$ part, we have the Knizhnik- Zamolodchikov relation%
\begin{equation}
\sigma^{a}J_{-1}^{a}|\alpha\rangle=(k+2)L_{-1}|\alpha\rangle
\end{equation}
Hence the contribution of the second term in (50) is the same as of the third
term. Replacing $L_{-1}$ by $\partial$ we derive the original equation of motion!

The group theory of the above is as following (this is the pattern to be
generalized below). With respect to left and right multiplication the matrix
$g$ transforms as $(\frac{1}{2},\frac{1}{2})\sim(0\oplus1)$ where we
decomposed with respect to the diagonal subgroup. As we act on $g$ with
$J\overline{J}$ we get an object $(\frac{1}{2}\otimes1,\frac{1}{2}\otimes
1$)$\sim(\frac{1}{2}\oplus\frac{3}{2},\frac{1}{2}\oplus\frac{3}{2}).$ The key
observation is that the $\frac{3}{2}$ part is a null vector, while the
$\frac{1}{2}$ part is a Virasoro descendant.

There is a parity odd relevant operator in the linear sigma model,
$O=\varepsilon_{\mu\nu}\overrightarrow{\phi}[\partial_{\mu}\overrightarrow
{\phi}\times\partial_{\nu}\overrightarrow{\phi]}$ . It becomes a topological
density in the limit when and if $\overrightarrow{\phi}^{2}\rightarrow const.$
Because of the fusion rules [14 ] , we conclude that in terms of the WZNW
fields we have $O\Rightarrow Tr(g).$ This is consistent with the fact that the
WZNW action is invariant under the simultaneous change of orientation,
$\varepsilon\Rightarrow-\varepsilon$ and $g\Rightarrow-g.$According to the
formula for the anomalous dimensions, $\Delta=\frac{j(j+1)}{k+2,}$ we see that
the dimension of the theta-term as well as the $n-$ field itself is equal to
$\frac{1}{4}$ (which is well known in the Haldane - Affleck approach). From
the linear sigma model point of view, this critical point requires fine-
tuning of two parameters, the mass term (temperature) and the theta term.

What happens at higher values of $k$ and for other groups ? Let us give some
non-trivial examples. Consider first $k=2$ and the matrix $g$ in the vector
representation, $g\sim(1,1)\sim(0\oplus1\oplus2).$ The current descendant
transforms as
\begin{equation}
J\overline{J}g\sim(0\oplus1\oplus2,0\oplus1\oplus2)\sim(0\oplus1,0\oplus1)
\end{equation}
$,$ since the formula (52 ) tells us that the spin two part is a null vector.
The formula (53) implies that the spin one object is a Virasoro descendent.
There are, however no constraints on the spin zero part. Therefore we can draw
the following conclusions. First, if we want to describe a linear sigma model
with the vector field $\overrightarrow{\phi}$ by WZNW action with $k=2,$ we
obtain the equation of motion with the non-vanishing Kac-Moody descendent
coming from $(1,0)+(0,1).$Generally this doesn't describe the conformal point
of the $\overrightarrow{n}$ field. But, if we fine tune the parameters to
eliminate this term, we do get a conformal sigma model. This conclusion is
consistent with the well known fact that while generally spin 1
antiferromagnet has a mass gap, there is a special, completely integrable
multicritical point, at which it is conformal (see [15 ] for a review).

This is not all, however. Consider a linear sigma model with the field of spin
2, described by a traceless symmetric tensor $\phi_{ij}.$We see from the (54 )
that in this case spin zero descendents do not appear . Hence we arrive at the
conclusion that spin 2 order parameter leads to a conformal theory without
extra fine tuning. The equations of motion in this case are%
\begin{equation}
\partial^{2}\phi_{ij}=\lambda\lbrack\phi^{3}]_{ij}%
\end{equation}
where the brackets mean the projection to spin 2 representation. We expect
this theory to be described by the free fermions.

It is easy to generalize these considerations for other order parameters and
symmetry groups. It might seem puzzling that we find conformal regimes in
cases, like $CP^{N}$ models, in which the large $N$ expansion tells us that
they don't exist. The resolution of the paradox lies in the fact that to reach
the above fixed points one has to add to the action operators containing
higher powers the fields and their derivatives. Such operators ( neglected and
intractable in the large $N$ expansion) might seem hopelessly irrelevant.
However, an old computation of their dimensions in [17 ] indicates that the
one loop correction tends to compensate the growth of the naive dimension. For
example, in the $O(3)$ case the operators in question have the form%
\begin{equation}
O_{sp}=(\partial_{z}n\partial_{\overline{z}}n)^{s}(\partial_{z}n)^{2p}%
(\partial_{\overline{z}}n)^{2p}%
\end{equation}
with the dimension (in the one loop approximation; $\alpha_{0}$ is a coupling
constant )%
\begin{equation}
\delta=s+2p+\alpha_{0}[2s(s-1)-16p]
\end{equation}
I believe that some of these operators become relevant at large p (as already
was conjectured in [16,17]) and drive the theories to higher fixed points.
They are identified with the finite number of relevant operators of WZNW at
level $k$ and must be fine-tuned at the conformal point.

These methods are easy to generalize to the case when the field $\phi$ is
defined via gauged WZNW theory. For example, the coset model%
\begin{equation}
M=\frac{SU(2)_{k}\times SU(2)_{1}}{SU(2)_{k+1}}%
\end{equation}
gives rise to the scalar field $\phi$ $=Tr$( $g_{1}^{-1}g_{2})$ $($where the
indices 1 and 2 refer to the factors in the numerator of (58 ) ) satisfying
the equation
\begin{equation}
\partial^{2}\phi=(\phi)^{2k+1}%
\end{equation}
in complete agreement with Zamolodchikov's results [13 ].

The last example in this section is the supergroup $OSp(1|2).$ This example is
important for analyses of the gauge/string dualities. The algebra $OSp(1|2)$
is obtained by adding to the three generators of $SU(2)$ , $J^{\pm},$ $J^{3}$,
two spinor anticommuting generators , $q^{\pm}$. The finite dimensional
representations are the pairs of symmetric spinors of ranks $j$ $=2s$ and
$j-1$ which transforms as%
\begin{align}
q_{\alpha}\Phi_{\alpha_{1}...\alpha_{j}}  &  =\sum\varepsilon_{\alpha
\alpha_{k}}\Psi_{\alpha_{1}...\widehat{\alpha}_{k}...\alpha_{j}}\\
q_{\alpha}\Psi_{\alpha_{1}...\alpha_{j-1}}  &  =\Phi_{\alpha\alpha
_{1}...\alpha_{j-1}}%
\end{align}
The dimension of these representations is $2s+1+2(s-\frac{1}{2})+1=4s+1.$ The
fundamental representation has dimension three and consists of a scalar and a
two components spinor. The matrix $g$ which enters the WZNW action is
3$\times$3. If we denote these representation as $[s],$they decompose into the
$SU(2)$ representations, denoted by the round brackets, as $[s]=(s)+(s-\frac
{1}{2}).$

We can identify the fields of the linear sigma model by the relations,
directly generalizing the $SU(2)$ case%
\begin{align}
\phi_{a}  &  =str(\sigma_{a}g)\\
\vartheta_{\alpha}  &  =str(\sigma_{\alpha}g)
\end{align}
where the $\sigma-s$ are the $3\times3$ generalization of the Pauli matrices.
The object we get transforms by the vector representation $[1]$ which has
$3+2$ dimensions. As before, we have to use the fusion rules to identify the
linear sigma model. Once again, we need to know the null vectors for
$OSp\{1|2).$All of them are known from the comprehensive work [18 ]. For our
purposes we need only the simplest ones which are easy to get directly. The
algebra has the form
\begin{align}
\lbrack J_{n}^{+},J_{m}^{-}]  &  =2J_{n+m}^{3}+nk\delta_{n+m,0}\\
\{q_{n}^{+},q_{m}^{-}\}  &  =2J_{n+m}^{3}+2nk\delta_{n+m,0}\\
\{q_{n}^{\pm},q_{m}^{\pm}\}  &  =\pm2J_{n+m}^{\pm}\\
\lbrack J_{n}^{\pm},q_{m}^{\mp}]  &  =-q_{n+m}^{\pm}%
\end{align}
while all other commutators are obvious. The null states at the first level
are as following%
\begin{align}
|f_{1}\rangle &  =J_{-1}^{+}|s\rangle\\
|f_{2}\rangle &  =(q_{-1}^{+}+J_{-1}^{+}q_{0}^{-})|s\rangle
\end{align}
where the highest weight state $|s\rangle$ satisfies the conditions $J_{0}%
^{+}|s\rangle=q_{0}^{+}|s\rangle=0,J_{0}^{3}|s\rangle=s|s\rangle.$ The above
states are null, provided that $s=\frac{k}{2}.$ Other null states at level 1
arise at $k=s-1,-2s-1,-s-\frac{3}{2}.$It is easy to write explicit expressions
for all of them. What we need, however , is a more invariant meaning of these
vectors. Namely, the multiplet of currents, $(J,q)$ transforms according to
$[1]$ representation of $OSp(1|2).$ When we act with the currents on the
operators in representation $[s],$ we have to decompose the product into
irreducible components. This is given by the formula%
\begin{equation}
\lbrack1]\times\lbrack s]=[s+1]+[s+\frac{1}{2}]+[s]+[s-\frac{1}{2}]+[s-1]
\end{equation}
(the only difference with the rotation group is that the half-integer spins
appear in the middle). It is easy to see now that at $s=\frac{k}{2}$ the whole
representation $[s+1]$ is formed by the null states (and thus drops out). At
$s=k+1$ the term $[s+\frac{1}{2}]$ is dumped. This pattern continues, removing
$[s-\frac{1}{2}]$ at $s=-k-\frac{3}{2}$ and $[s-1]$ at $s=-\frac{k+1}{2}.$

The null vectors are important here for two reasons. First, they define the
fusion rules. For example, a simple use of the Ward identities show that at
$k=1$ the primary field $\varphi$ of the representation $[\frac{1}{2}]$
satisfies the fusion%
\begin{equation}
\varphi\times\varphi=I+\varphi
\end{equation}
where $I$ is a unit operator ; the difference with the $SU(2)$ fusions is in
the second term . More over, $\varphi\times\varphi\times\varphi=I+2\varphi,$
and as a result the $OSp$ models are very different from their $O(3)$
counterparts. A naive idea to describe the theory as a linear sigma model with
the interaction $V\sim(\overrightarrow{\phi}^{2}+\overline{\vartheta}%
\vartheta$ )$^{2}$ fails because the equations of motion will contain
descendants of the unit operator (proportional to the product of the two
currents). This is just as well, since a model like that would have larger
symmetry $OSp$ $(3|2)$ (reflecting the absence of the spin-orbit interaction)
which we certainly don't need . In order to get a conformal theory we have to
combine the $\varphi^{4}$ and $\varphi^{3}$ terms in such a way as to cancel
the unit operator in the equations of motion. The general structure of these
cubic terms is
\begin{equation}
L_{3}=c_{1}(\overline{\vartheta}\gamma^{a}\partial_{\mu}\vartheta
)\partial_{\mu}\phi^{a}+c_{2}\varepsilon_{\mu\nu}(\overline{\vartheta}%
\gamma^{a}\partial_{\mu}\vartheta)\partial_{\nu}\phi^{a}%
\end{equation}
These terms create the spin-orbit interaction and are analogous to the terms
in the Green-Schwartz action. We should notice that this model

necessarily contains the negative norms. Coupling to gravity and imposing
$\kappa-$ symmetry should eliminate them but the concrete mechanisms of this
are not clear at present. Perhaps once again this linear sigma model is
equivalent to the (45) but it is not proved.

\section{ Using connections}

Let us describe yet another approach to the same class of problems. Let us
look again at the $O(3)$ invariant $n$ -field, describing it this time in
terms of the $SU(2)$ connections , which we denote as $(B_{\mu}^{a}$ ,$A_{\mu
})$ where $a=1,2$ and the direction 3 is selected for the gauge group
$U(1).$These connections satisfy the Maurer- Cartan equations (the zero field
strength condition)%
\begin{align}
\nabla_{+}B_{-}-\nabla_{-}B_{+}  &  =0\\
\partial_{-}A_{+}-\partial_{+}A_{-}  &  =\epsilon^{ab}B_{-}^{a}B_{+}^{b}%
\end{align}
where $\nabla_{+}B^{a}=\partial_{+}B^{a}+\epsilon^{ab}A_{+}B^{b}$ is the
$O(2)$ covariant derivative. The invariant lagrangian has the form%
\begin{equation}
L=\frac{1}{2\alpha}B_{+}^{a}B_{-}^{a}+\theta\epsilon^{ab}B_{+}^{a}B_{-}^{b}%
\end{equation}
and the equations of motion are%
\begin{equation}
\nabla_{+}B_{-}+\nabla_{-}B_{+}=0
\end{equation}
Let us compare them with the equations for the WZNW action for $SU(2),$written
in terms of the same three connections $(A,B).$Of course the first pair of the
equations ( the zero field strength ) is precisely the same as (73,74 ). But
the equations of motion this time are
\begin{equation}
\partial_{-}A_{+}=\partial_{-}B_{+}=0
\end{equation}
meaning that we have three holomorphic currents. Unlike (76) there is no gauge
symmetry in these equations.

To make a comparison of the two models let us chose a gauge $A_{-}=0$ for the
$n$- field. The equations of motion take the form%
\begin{align}
\partial_{-}B_{+} &  =0\\
\partial_{-}A_{+} &  =\epsilon^{ab}B_{+}^{a}B_{-}^{b}%
\end{align}
and are , generally speaking, different from (77 ).

However, if we chose $k=1$ in WZNW , something special happens. Let us
remember that in this theory there are two sets of currents. The first one (
which we have denoted $A_{+},B_{+}$ ) are defined as $J_{+}=g^{-1}\partial
_{+}g$ and $J_{-}=g\partial_{-}g^{-1}$ are holomorphic and have normal
dimensions. Let us call them "direct currents" (DC). The second set consists
of the " alternative currents" (AC), which are $K_{+}=g\partial_{+}g^{-1}$ and
$K_{-}=g^{-1}\partial_{-}g.$ For a general WZNW theory the alternative
currents do not conserve and acquire the anomalous dimension $\Delta
=1+\frac{b}{k+b},$ where $b$ is the dual Coxeter number ( adjoint Casimir).
Notice that $K_{+}=-g^{-1}J_{+}g.$ That implies that AC are the current
descendants of the adjoint operator $g^{-1}\otimes g.$

The crucial point for us is that in the equation ( 79) the $A_{+}$ and $B_{+}$
are the components of the direct current while the $B_{-}$ is an AC. As we
already discussed in the last section, at $k=1$ the adjoint operator decouples
from the OPE. Hence, at this value of $k$ the $B_{-}$ field can be set to zero
and the WZNW equations become identical to the $n-$ field equations! Thus, as
expected, the $n$ -field theory has a conformal fixed point equivalent to the
WZNW fixed point at $k=1$. Now, let us take $k>1.$This time the $B_{-}$ terms
don't disappear by themselves. They must be forced out by adding the higher
derivative operators ( 56) and fine-tuning their couplings. This explains why
the integrable antiferromagnets of higher spin $S$ are described by the
$k=2S$  theory. It is straightforward to generalize this method for the
general case $G/H.$ It requires, however a careful analysis of the null vectors.

More conformal points arise when we use gauged WZNW to solve the coset theory,
" cosets for cosets". Let us describe the general idea, leaving the details
for the future. As is well known, gauging of WZNW refers to the diagonal
group, $g\sim h^{-1}gh,$ while the cosets we are interested in are always
$g\sim gh.$ Surprisingly there is a connection between the two.  In general we
should compare the diagonal coset $G/F$ with the right coset $G/H,$ but here
we restrict ourselves  with the simplest case of $SU(2)/U(1).$We will also
write the gauged action in a somewhat non-standard form%
\begin{equation}
S=W_{G}[B_{+},A_{+}]-W_{H}[A_{+}]
\end{equation}
where $W$-s are the WZNW actions for the groups $G$ and $H$ as functionals of
the left-invariant connections. Notice that in this definition both $A$ and
$B$ are defined for the group $G$ (and this $A$ is substituted to the action
for the group $H$ ). By the use of the standard variation formulae [19 ] it is
easy to obtain the equations of motion%
\begin{align}
\partial_{-}A_{+}-\partial_{+}A_{-} &  =[B_{+}B_{-}]\\
\nabla_{-}B_{+}-\nabla_{+}B_{-} &  =0\\
\widetilde{\nabla}_{-}B_{+} &  =0
\end{align}
where $\nabla=\partial+A$ and $\widetilde{\nabla}$=$\partial+\widetilde{A}$
with the definition $\partial_{+}\widetilde{A}_{-}-\nabla_{-}A_{+}=0.$These
equations have obvious $H$ gauge invariance. In the gauge $A_{+}=0$ they
coincide with the standard gauged WZNW in the gauge $a_{+}=0$ (where $a$ is a
gauge field). Indeed in the latter case we have an action
\begin{equation}
S=W[g]+\int a_{-}A_{+}%
\end{equation}
Integration over $a$ enforces $A_{+}=0$ condition and the extra term leads to
the above equations with $\widetilde{A}_{-}=a_{-}.$

Once again by exploiting various null vectors at various $k$ one can "solve"
the local $G/H$ theory by the gauged WZNW. We shall leave it for the future work.

\section{Supertwistors and zero curvature}

In the ordinary sigma models one of the ways of finding the Bethe ansatz is to
reformulate the model in terms of the multi-flavored fermions [19 ]. It uses
the fact that as the number of flavors go to infinity, the fermions in the
external gauge fields impose the condition that the field strength is zero.
For the lagrangians quadratic in the gauge fields they can be integrated out,
giving the four-fermion interaction. For such models the Bethe ansatz is known
and all one had to do to describe the sigma model was to take the limit of
infinite flavors in the Bethe equations. In this section we will outline the
first part of this procedure - reduction to the four-particle interaction.

Let us discuss again the case of $\frac{OSp(1|2)}{0(2)}$ (other cases are
similar). The simplest representation which we shall use is a supertwistor
$\Lambda_{\pm}=$ $(\lambda_{\pm}^{\alpha},\phi_{\pm})$ which consists of the
bosonic spinor $\lambda$ (with two Majorana components) and fermionic scalar
$\phi,$ which are spinors on the world sheet. The free lagrangian has the form%
\begin{equation}
L=\overline{\lambda}\widehat{\partial}\lambda+\phi\widehat{\partial}%
\phi=\overline{\Lambda}\widehat{\partial}\Lambda
\end{equation}
where $\widehat{\partial}$ is a two dimensional Dirac operator and as usual
$\overline{\lambda}^{\alpha}=\epsilon^{\alpha\beta}\lambda^{\beta}$ . Notice
that the inverted statistics of these fields is a must - otherwise the
lagrangian is a total derivative. There are five currents of $OSp(1|2)$ -
three of $SL(2),$ $\overline{\lambda}\tau^{i}\lambda$ and two spinor currents
$\lambda^{\alpha}\phi.$ They couple to the five gauge fields $(A,B^{a}%
,\psi^{\alpha}).$The covariant derivatives have the form%
\begin{align}
\nabla\lambda &  =\partial\lambda+B^{a}\tau^{a}\lambda+A\tau^{3}\lambda
+\psi\phi\\
\nabla\phi &  =\partial\phi+\overline{\psi}\lambda
\end{align}
When we consider an infinite number of copies of these twistors and integrate
them out, we get, as was shown in [19 ] a delta function of the field
strength. Hence, if we add the lagrangian ( 85) to (45 ), in the infinite
flavour limit we obtain the desired sigma model. On the other hand we can
integrate out the connections first and get the model with quartic interaction
of supertwistors which has the form%
\begin{equation}
L_{int}\sim(\phi_{+}\phi_{-})(\overline{\lambda}_{+}\tau^{3}\lambda
_{-})+(\overline{\lambda}_{+}\tau^{a}\lambda_{\_})^{2}%
\end{equation}
This interaction is presumably completely integrable and should produce the
$R$ matrix for supertwistors satisfying the Yang-Baxter equations. This matrix
should be of XXZ (trigonometric) type and perhaps coincide with the already
known ones [8]. If so , finding the Bethe equations should be relatively
straightforward. There is a slight complication because of the $O(2)$ gauge
symmetry. In the bosonic case this difficulty can be overcome by adding the
term $A^{2}$ to the action (thus getting the $XXZ$ model) and then sending its
coefficient to zero. The same should work in the present case.

\section{Conclusions}

The present work is a first step towards finding CFT-s describing sigma
models. I think that it will require a work of many people to complete this
task. I also believe that if we want to understand fundamental physics, there
is no way avoiding it. The nature of space- time at large curvature is
determined by gauge /strings correspondence and it , to the large extent,
boils down to the sigma models, similar to the ones we discussed. It is
possible that when better understood these ideas will influence our picture of
the early universe.

Other important physical problems involving sigma models are the quantum Hall
effect and the theory of 2d turbulence (important for meteorology). In both
cases we have to deal with a non-compact CFT. In the case of turbulence there
are general considerations [21]  and recent interesting numerical data [22],
supporting this view.

On the other hand it should be said that the problem is very difficult. It
will require considerable amplification of the already esoteric technics of
Bethe ansatz and/or some intricate analysis of the CFT algebras. Also,it is
possible that some important fixed points are of the different nature than the
ones described in this paper. Still I am hopeful that these difficulties will
be resolved.

I would like to thank my friends, B. Altshuler, N. Berkovits,  I. Klebanov, N.
Nekrasov, L. Takhtajan ,P. Wiegmann and A. Zamolodchikov for very helpful discussions.

This work was partially supported by the NSF grant 0243680. Any opinions,
findings and conclusions or recommendations expressed in this material are
those of the authors and do not necessarly reflect the views of the National
Science Foundation.

\bigskip

REFERENCES

[1] A.M. Polyakov Phys. Lett.B,59 (1975) 85

[2] F. Haldane Phys. Rev. Lett. 50 (1983) 1153

[3] I. Affleck Journ.of Phys. of Cond. Matt. 1(1989)3047

[4] E. Witten Nucl. Phys.B223 (1983) 4222

[5] L. Faddeev, G. Korchemsky Phys. Lett.B 342(1995)311, G Korchemsky Nucl.
Phys. B550 (1999) 397

[6] L. Takhtajan Phys. Lett. A 87 (1982) 479

[7] H. Babujan Nucl. Phys. B215 (1983) 317

[8] P. Kulish J. Sov. Math. 35 (1986) 1111

[9] A. Polyakov Int. Journ. Mod Phys. A14 (1999)645, hep-th/9809057

[10] H. Verlinde PUPT/2113, hep-th/0403024

[11] A. Polyakov Mod. Phys. Lett. 19 (2004) 22 hep-th/

[12] A. Polyakov ZhETF (sov. phys.) 55 (1968)1026

[13] A. Zamolodchikov Sov. Journ. Nucl.Phys. 44 (1986) 529

[14] V. Knizhnik, A. Zamolodchikov Nucl. Phys. B247(1984) 83

[15] L. Faddeev N. Reshetikhin Ann. Phys. 167 (1986) 363

[16] V. Kravtsov I. Lerner V. Yudson Phys. Lett. A 134 (1989)245

[17] F. Wegner Z. Phys. B78 (1990) 33

[18] V. Kac Y. Wakimoto Proc. Nat. Acad. USA 85 (1986)4956

[19] A. Polyakov P. Wiegmann Phys. Lett b 131 (1983) 121

[20] N. Berkovits et al. Nucl. Phys. B567 (2000)61 , hep-th/9907200

[21] A. Polyakov Nucl. Phys. B396 (1993) 367

[22] D. Bernard et al. Nature (in print)

\end{document}